\errorstopmode
\input amssym.def
\input amssym.tex


\magnification=\magstephalf
\hsize=14.0 true cm
\vsize=19 true cm
\hoffset=1.0 true cm
\voffset=2.0 true cm

\abovedisplayskip=12pt plus 3pt minus 3pt
\belowdisplayskip=12pt plus 3pt minus 3pt
\parindent=1.0em


\font\sixrm=cmr6
\font\eightrm=cmr8
\font\ninerm=cmr9

\font\sixi=cmmi6
\font\eighti=cmmi8
\font\ninei=cmmi9

\font\sixsy=cmsy6
\font\eightsy=cmsy8
\font\ninesy=cmsy9

\font\sixbf=cmbx6
\font\eightbf=cmbx8
\font\ninebf=cmbx9

\font\eightit=cmti8
\font\nineit=cmti9

\font\eightsl=cmsl8
\font\ninesl=cmsl9

\font\sixss=cmss8 at 8 true pt
\font\sevenss=cmss9 at 9 true pt
\font\eightss=cmss8
\font\niness=cmss9
\font\tenss=cmss10

 at 12 true pt
 at 12 true pt
\font\bigrm=cmr10 at 12 true pt
 at 12 true pt
 at 12 true pt

 at 16 true pt
 at 16 true pt
\font\Bigrm=cmr12 at 16 true pt
 at 16 true pt
 at 16 true pt

\catcode`@=11
\newfam\ssfam

\def\tenpoint{\def\rm{\fam0\tenrm}%
    \textfont0=\tenrm \scriptfont0=\sevenrm \scriptscriptfont0=\fiverm
    \textfont1=\teni  \scriptfont1=\seveni  \scriptscriptfont1=\fivei
    \textfont2=\tensy \scriptfont2=\sevensy \scriptscriptfont2=\fivesy
    \textfont3=\tenex \scriptfont3=\tenex   \scriptscriptfont3=\tenex
    \textfont\itfam=\tenit                  \def\it{\fam\itfam\tenit}%
    \textfont\slfam=\tensl                  \def\sl{\fam\slfam\tensl}%
    \textfont\bffam=\tenbf \scriptfont\bffam=\sevenbf
    \scriptscriptfont\bffam=\fivebf
                                            \def\bf{\fam\bffam\tenbf}%
    \textfont\ssfam=\tenss \scriptfont\ssfam=\sevenss
    \scriptscriptfont\ssfam=\sevenss
                                            \def\ss{\fam\ssfam\tenss}%
    \normalbaselineskip=13pt
    \setbox\strutbox=\hbox{\vrule height8.5pt depth3.5pt width0pt}%
    \let\big=\tenbig
    \normalbaselines\rm}

\def\ninepoint{\def\rm{\fam0\ninerm}%
    \textfont0=\ninerm      \scriptfont0=\sixrm
                            \scriptscriptfont0=\fiverm
    \textfont1=\ninei       \scriptfont1=\sixi
                            \scriptscriptfont1=\fivei
    \textfont2=\ninesy      \scriptfont2=\sixsy
                            \scriptscriptfont2=\fivesy
    \textfont3=\tenex       \scriptfont3=\tenex
                            \scriptscriptfont3=\tenex
    \textfont\itfam=\nineit \def\it{\fam\itfam\nineit}%
    \textfont\slfam=\ninesl \def\sl{\fam\slfam\ninesl}%
    \textfont\bffam=\ninebf \scriptfont\bffam=\sixbf
                            \scriptscriptfont\bffam=\fivebf
                            \def\bf{\fam\bffam\ninebf}%
    \textfont\ssfam=\niness \scriptfont\ssfam=\sixss
                            \scriptscriptfont\ssfam=\sixss
                            \def\ss{\fam\ssfam\niness}%
    \normalbaselineskip=12pt
    \setbox\strutbox=\hbox{\vrule height8.0pt depth3.0pt width0pt}%
    \let\big=\ninebig
    \normalbaselines\rm}

\def\eightpoint{\def\rm{\fam0\eightrm}%
    \textfont0=\eightrm      \scriptfont0=\sixrm
                             \scriptscriptfont0=\fiverm
    \textfont1=\eighti       \scriptfont1=\sixi
                             \scriptscriptfont1=\fivei
    \textfont2=\eightsy      \scriptfont2=\sixsy
                             \scriptscriptfont2=\fivesy
    \textfont3=\tenex        \scriptfont3=\tenex
                             \scriptscriptfont3=\tenex
    \textfont\itfam=\eightit \def\it{\fam\itfam\eightit}%
    \textfont\slfam=\eightsl \def\sl{\fam\slfam\eightsl}%
    \textfont\bffam=\eightbf \scriptfont\bffam=\sixbf
                             \scriptscriptfont\bffam=\fivebf
                             \def\bf{\fam\bffam\eightbf}%
    \textfont\ssfam=\eightss \scriptfont\ssfam=\sixss
                             \scriptscriptfont\ssfam=\sixss
                             \def\ss{\fam\ssfam\eightss}%
    \normalbaselineskip=10pt
    \setbox\strutbox=\hbox{\vrule height7.0pt depth2.0pt width0pt}%
    \let\big=\eightbig
    \normalbaselines\rm}

\def\tenbig#1{{\hbox{$\left#1\vbox to8.5pt{}\right.\n@space$}}}
\def\ninebig#1{{\hbox{$\textfont0=\tenrm\textfont2=\tensy
                       \left#1\vbox to7.25pt{}\right.\n@space$}}}
\def\eightbig#1{{\hbox{$\textfont0=\ninerm\textfont2=\ninesy
                       \left#1\vbox to6.5pt{}\right.\n@space$}}}

\font\sectionfont=cmbx10
\font\subsectionfont=cmti10

\def\figurecaptionfont{\ninepoint}
\def\tablecaptionfont{\ninepoint}
\def\footnotefont{\eightpoint}


\newcount\equationno
\newcount\bibitemno
\newcount\figureno
\newcount\tableno

\equationno=0
\bibitemno=0
\figureno=0
\tableno=0


\footline={\ifnum\pageno=0{\hfil}\else
{\hss\rm\the\pageno\hss}\fi}


\def\section #1. #2 \par
{\vskip0pt plus .10\vsize\penalty-100 \vskip0pt plus-.10\vsize
\vskip 1.6 true cm plus 0.2 true cm minus 0.2 true cm
\global\def\equationlabel{#1}
\global\equationno=0
\leftline{\sectionfont #1. #2}\par
\immediate\write\terminal{Section #1. #2}
\vskip 0.7 true cm plus 0.1 true cm minus 0.1 true cm
\noindent}


\def\subsection #1 \par
{\vskip0pt plus 0.8 true cm\penalty-50 \vskip0pt plus-0.8 true cm
\vskip2.5ex plus 0.1ex minus 0.1ex
\leftline{\subsectionfont #1}\par
\immediate\write\terminal{Subsection #1}
\vskip1.0ex plus 0.1ex minus 0.1ex
\noindent}


\def\appendix #1. #2 \par
{\vskip0pt plus .20\vsize\penalty-100 \vskip0pt plus-.20\vsize
\vskip 1.6 true cm plus 0.2 true cm minus 0.2 true cm
\global\def\equationlabel{\hbox{\rm#1}}
\global\equationno=0
\leftline{\sectionfont Appendix #1. #2}\par
\immediate\write\terminal{Appendix #1. #2}
\vskip 0.7 true cm plus 0.1 true cm minus 0.1 true cm
\noindent}



\def\equation#1{$$\displaylines{\qquad #1}$$}
\def\enum{\global\advance\equationno by 1
\hfill\llap{{\rm(\equationlabel.\the\equationno)}}}

\def\next#1{\cr\noalign{\vskip#1}\qquad}


\def\ifundefined#1{\expandafter\ifx\csname#1\endcsname\relax}

\def\ref#1{\ifundefined{#1}?\immediate\write\terminal{unknown reference
on page \the\pageno}\else\csname#1\endcsname\fi}

\newwrite\terminal
\newwrite\bibitemlist

\def\bibitem#1#2\par{\global\advance\bibitemno by 1
\immediate\write\bibitemlist{\string\def
\expandafter\string\csname#1\endcsname
{\the\bibitemno}}
\item{[\the\bibitemno]}#2\par}

\def\beginbibliography{
\vskip0pt plus .15\vsize\penalty-100 \vskip0pt plus-.15\vsize
\vskip 1.2 true cm plus 0.2 true cm minus 0.2 true cm
\leftline{\sectionfont References}\par
\immediate\write\terminal{References}
\immediate\openout\bibitemlist=biblist
\frenchspacing\parindent=1.8em
\vskip 0.5 true cm plus 0.1 true cm minus 0.1 true cm}

\def\endbibliography{
\immediate\closeout\bibitemlist
\nonfrenchspacing\parindent=1.0em}


\def\figurecaption#1{\global\advance\figureno by 1
\narrower\figurecaptionfont
Fig.~\the\figureno. #1}

\def\tablecaption#1{\global\advance\tableno by 1
\vbox to 0.25 true cm { }
\centerline{\tablecaptionfont%
Table~\the\tableno. #1}
\vskip-0.4 true cm}

\def\thicktablerule{\hrule height1pt}
\def\thintablerule{\hrule height0.4pt}

\tenpoint

\immediate\openin\bibitemlist=biblist
\ifeof\bibitemlist\immediate\closein\bibitemlist
\else\immediate\closein\bibitemlist
\input biblist \fi


\def\thismonth{\ifcase\month\or
January\or February\or March\or April\or May\or June\or
July\or August\or September\or October\or November\or December\fi}

\input epsf
\epsfclipon



\def\rmd{{\rm d}}

\def\rmO{{\rm O}}
\def\rmU{{\rm U}}



\def\proof{\noindent{\sl Proof:}\kern0.6em}

\def\frac#1#2{\hbox{$#1\over#2$}}
\def\dual{\mathstrut^*\kern-0.1em}

\def\lvec#1{\setbox0=\hbox{$#1$}
    \setbox1=\hbox{$\scriptstyle\leftarrow$}
    #1\kern-\wd0\smash{
    \raise\ht0\hbox{$\raise1pt\hbox{$\scriptstyle\leftarrow$}$}}
    \kern-\wd1\kern\wd0}
\def\rvec#1{\setbox0=\hbox{$#1$}
    \setbox1=\hbox{$\scriptstyle\rightarrow$}
    #1\kern-\wd0\smash{
    \raise\ht0\hbox{$\raise1pt\hbox{$\scriptstyle\rightarrow$}$}}
    \kern-\wd1\kern\wd0}
\def\slash#1{\setbox0=\hbox{$#1$}\setbox1=\hbox{$\kern1pt/$}
    #1\kern-\wd0\kern1pt/\kern-\wd1\kern\wd0}


\def\nab#1{{\nabla_{#1}}}
\def\nabstar#1{{\nabla\kern0.5pt\smash{\raise 4.5pt\hbox{$\ast$}}
               \kern-5.5pt_{#1}}}

\def\drvstar#1{{\partial\kern0.5pt\smash{\raise 4.5pt\hbox{$\ast$}}
               \kern-6.0pt_{#1}}}

\def\ldrvstar#1{{\lvec{\,\partial}\kern-0.5pt\smash{\raise 4.5pt\hbox{$\ast$}}
               \kern-5.0pt_{#1}}}


\def\MeV{{\rm MeV}}

\def\fm{{\rm fm}}



\def\psibar{\overline{\psi}}


\def\dirac#1{\gamma_{#1}}
\def\diracstar#1#2{
    \setbox0=\hbox{$\gamma$}\setbox1=\hbox{$\gamma_{#1}$}
    \gamma_{#1}\kern-\wd1\kern\wd0
    \smash{\raise4.5pt\hbox{$\scriptstyle#2$}}}


\def\Ad{{\rm Ad}\kern0.1em}


\def\Dw{D_{\rm w}}
\def\Dm{D_{\rm m}}
\def\hDm{Q_{\rm m}}

\def\gap{\mu}
\def\gapbar{\bar{\mu}}

\def\thresh{\bar{\alpha}}
\def\threshR{\thresh_{\hbox{\sixrm R}}}

\def\ZA{Z_{\rm A}}
\def\ZP{Z_{\rm P}}
\def\ZS{Z_{\rm S}}
\def\rhoR{\rho_{\hbox{\sixrm R}}}
\def\csw{c_{\rm sw}}
\def\Ncfg{N_{\rm cfg}}

\def\bp{b_{\rm P}}

\def\bpp{b_{\rm P\kern-0.4pt P}}

\def\mq{m_{\rm q}}
\def\mr{m_{{\hbox{\sixrm R}}}}

\def\mpi{m_{\pi}}

%
\rightline{CERN-PH-TH/2005-246}

\vskip 1.5cm 
\centerline{\Bigrm Stability of lattice QCD simulations and}
\vskip 0.3cm
\centerline{\Bigrm the thermodynamic limit}
\vskip 0.6cm
\centerline{\bigrm L.~Del Debbio$^{1}$, L.~Giusti$^{1,}$%
\footnote{$^{\ast}$}{\footnotefont%
On leave from
Centre Physique Th\'eorique, CNRS Luminy, F-13288 Marseille, France},
M.~L\"uscher$^1$, R.~Petronzio$^2$, N.~Tantalo$^{2,3}$}
\vskip2.0ex
\centerline{$^1\hskip-3pt$ \it
CERN, Physics Department, TH Division, CH-1211 Geneva 23, Switzerland}
\vskip1.0ex
\centerline{$^2\hskip-3pt$ \it
Universit\`a di Roma ``Tor Vergata'' and INFN sezione ``Tor Vergata'',}
\centerline{\it Via della Ricerca Scientifica 1,  I-00133 Rome, Italy}
\vskip1.0ex
\centerline{$^3\hskip-3pt$ \it
Centro Enrico Fermi, Via Panisperna 89 A, I-00184 Rome, Italy}
\vskip 0.8 true cm
\thintablerule
\vskip 2.0ex
\ninepoint
\leftline{\bf Abstract}
\vskip 1.0ex\noindent
We study the spectral gap of the Wilson--Dirac operator
in two-flavour lattice QCD as a function of the lattice spacing $a$,
the space-time volume $V$ and the current-quark mass $m$.
It turns out that the median
of the probability distribution of the gap scales proportionally to $m$
and that its width is practically equal to
$a/\sqrt{V}$. In particular,
numerical simulations are safe 
from accidental zero modes
in the large-volume regime of QCD.
\vskip 2.0ex
\thintablerule

\tenpoint
\vskip-0.2cm

\section 1. Introduction

The formulation of lattice QCD proposed by Wilson long ago 
preserves many symmetries of the continuum theory
exactly [\ref{Wilson}].
An infamous exception are the chiral symmetries, and
although the symmetry-violating terms 
vanish proportionally to the lattice spacing
(or its square if the theory is Symanzik-improved [\ref{SW},\ref{OaImp}]),
the presence of these terms complicates the lattice theory considerably.
In particular, the fact that the massive Wilson--Dirac operator is not
protected from arbitrarily small eigenvalues
may lead to instabilities in numerical simulations.

Such instabilities were not observed, however,
in recent simulations of the two-flavour Wilson theory on large lattices,
not even at the smallest quark masses considered 
[\ref{SchwarzAlgorithmII}--\ref{NextPaper}].
It is tempting to attribute the absence of instabilities
in these simulations
to the use of a new simulation algorithm, but
this explanation cannot be right, because
the distribution of the spectral gap of the lattice Dirac 
operator (and thus the probability to find exceptionally small
eigenvalues) is determined by the lattice action and the functional
integral, and not by the method used to evaluate the latter.

Our aim in this paper is to clarify the situation
by calculating the distribution of the gap on 
a set of lattices, using numerical simulations.
In particular, we wish to determine, as explicitly as possible, 
the range in parameter space, where the Wilson theory can 
be simulated without running into instabilities.

\section 2. Algorithmic stability and the spectral gap

In this section we introduce our notation and discuss
the relevance of the distribution of the spectral gap
for the stability of lattice QCD simulations. For any unexplained
notation see ref.~[\ref{OaImp}].

\subsection 2.1 Lattice Dirac operator

The lattice theory is set up as usual on
hypercubic lattices with spacing $a$. 
Periodic boundary conditions are
imposed on all fields and in all directions,
except for the quark fields, which are taken to be antiperiodic in time.
Throughout this paper, we assume that 
there is a doublet of sea quarks with equal mass, although
many results are likely to remain valid if the number of 
quark flavours is larger than two.

While the action of the gauge field will always be the 
Wilson plaquette action, we shall consider various lattice 
Dirac operators $\Dm$ (the subscript indicates that
$\Dm$ includes the quark mass term). The Wilson--Dirac operator
\equation{
  \Dm=\Dw+m_0,
  \enum
  \next{2ex}
  \Dw=\frac{1}{2}\left\{\dirac{\mu}\left(\nabstar{\mu}+\nab{\mu}\right)
  -a\nabstar{\mu}\nab{\mu}\right\},
  \enum
}
is the one we are primarily interested in, as well as its 
O($a$)-improved version [\ref{SW},\ref{OaImp}].
In these equations, $\nab{\gap}$ and $\nabstar{\gap}$ denote the 
gauge-covariant forward and backward difference operators, and $m_0$
the bare quark mass. 

Rather than the Dirac operator itself, we prefer to consider
the hermitian operator
\equation{
  \hDm=\dirac{5}\Dm
  \enum
}
in the following. The determinants of these two operators are the same,
but the fact that the spectrum of $\hDm$ is real simplifies the discussion
considerably. On a finite lattice, and for any specified gauge field,
we then define the spectral gap
\equation{
  \gap=\min
  \left\{|\lambda|\bigm|\hbox{%
  $\lambda$ is an eigenvalue of $\hDm$}\right\}.
  \enum
}
Evidently, the gap is a well-defined function of the gauge field
and so is the spectral asymmetry
\equation{
  \eta=\frac{1}{2}\left\{N_{+}-N_{-}\right\},
  \enum
}
where $N_{\pm}$
are the numbers of positive and negative eigenvalues of $\hDm$.

In formulations of lattice QCD that preserve chiral symmetry
via the Ginsparg--Wilson relation [\ref{GinspargWilson}--\ref{ExactChSy}], 
the gap is bounded from below by the bare current-quark mass $m$
and the asymmetry vanishes if $m>0$ 
(see ref.~[\ref{FerencReview}], for example).
However, these properties are not guaranteed in the Wilson theory,
and it is possible, in the presence of some 
gauge-field configurations, that the gap is much
smaller than $m$ and that the asymmetry
assumes a non-zero value.

\subsection 2.2 Sources of instability

In the large-volume regime of QCD, and at large quark masses, 
the probability distribution $p(\mu)$ of the gap
typically looks like the one
shown in the upper plot in fig.~1. The subsets of gauge fields,
where $\mu$ is 
far below the central value of the distribution, occur with such a small 
probability in this case that their contributions
to the common physical observables can be safely neglected.
Numerical simulations, using a 
preconditioned Hybrid Monte Carlo (HMC) algorithm [\ref{HMC}], for example, 
will then normally run smoothly and produce a representative sample 
of field configurations as expected.

\topinsert
\vbox{
\vskip0.0cm
\epsfxsize=7.0cm\hskip2.5cm\epsfbox{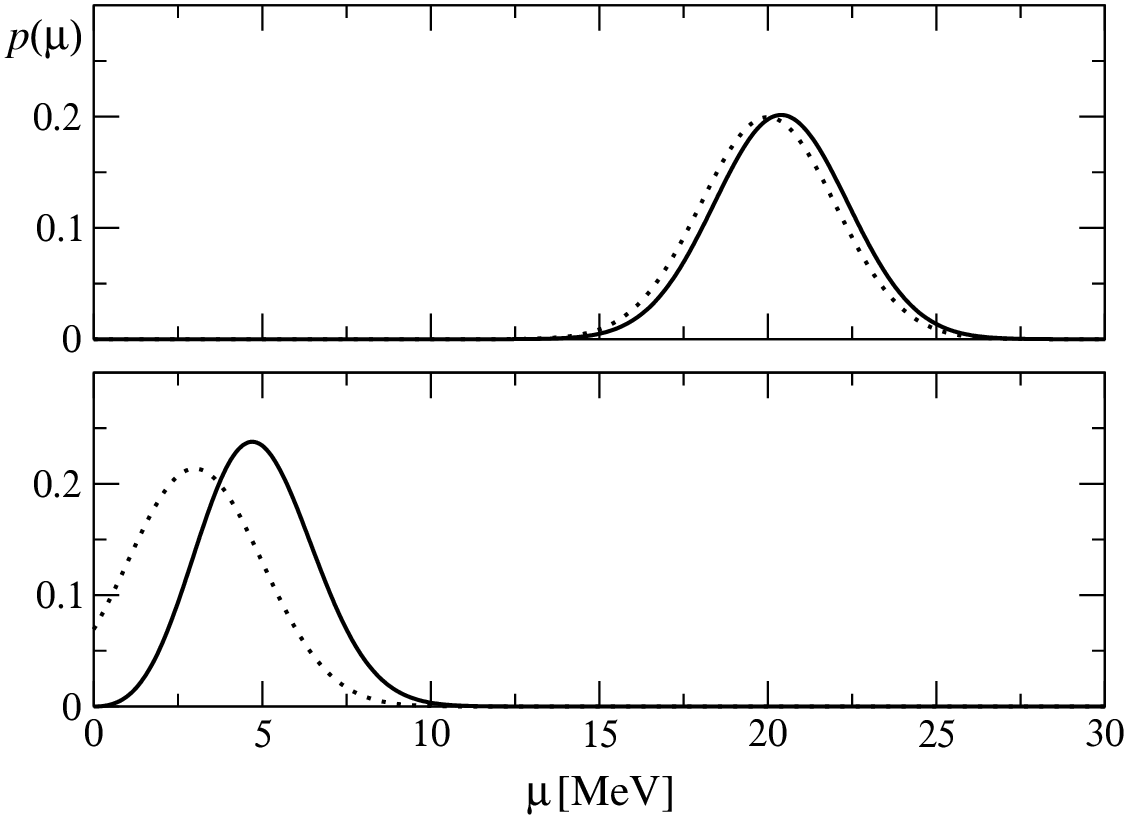}
\vskip0.3cm
\figurecaption{%
Qualitative form of the normalized probability distribution $p(\mu)$ of the gap
(full lines), at two values of the quark mass (upper and lower plot).
The normalized weighted distributions 
proportional to $p(\mu)/\mu^2$ are also shown (dotted lines).
}
\vskip0.0cm
}
\endinsert

When the quark mass decreases, 
the centre of the gap distribution moves towards 
smaller values and eventually is no more than
one or two standard deviations away from the origin
(lower plot in fig.~1). At this point
numerical simulations may run into instabilities for the following reasons:

\vskip1.5ex
\noindent
(1) {\it Integration instabilities}.
The HMC algorithm obtains the next field configuration by
integrating the appropriate molecular-dynamics equations in 
field space, followed by an accept--reject step at the end of the 
integration. It is possible that
the molecular-dynamics trajectories pass through field configurations
where the gap $\mu$ is exceptionally small.
The numerical integration 
becomes unstable in this case and liable to rounding errors.
In particular, the reversibility of the integration is then no longer
guaranteed, which invalidates the algorithm.

\vskip1ex
\noindent
(2) {\it Ergodicity}. Along the molecular-dynamics trajectories, the 
spectral asymmetry $\eta$ tends to be preserved. Changes in the 
asymmetry are in fact excluded in the limit where 
the integration step-size goes to zero. 
However, the sectors in field space, in which the asymmetry
does not vanish, may become statistically relevant
in the situation considered here.
The HMC algorithm is then likely to give wrong results, because
it may get stuck, for a very long time, in any one of these sectors. 

\vskip1ex
\noindent
(3) {\it Sampling inefficiencies}.
Observables that are sensitive to the small eigenvalues of 
the Dirac operator, such as the pion propagator, 
may be poorly sampled by the simulation.
The effect is illustrated
by the dotted curves in fig.~1, which represent the distributions
of the gap
reweighted by the ``observable'' $1/\mu^2$. 
Expectation values of such quantities are difficult to compute
reliably, since the
subspace of fields where $\mu$ is very small is only 
rarely visited in the course of the simulation.

\vskip1.5ex
\noindent
Instabilities of this kind can lead to underestimated
statistical errors and incorrect results, and they may even suggest the 
presence of a phase transition when there is none. 
Since the QCD functional integral remains well defined, also in this 
difficult regime, improved simulation techniques 
may conceivably be developed, which do not 
suffer from any instabilities. 
Whether this is worth the effort is not obvious, however,
because the theory may be strongly affected by lattice artefacts
(resulting from a competition of mass and discretization effects)
in the critical range of parameters.

\section 3. Numerical studies

In the presence of any given gauge field,
the spectral gap of the Dirac operator 
can be computed numerically,
using a suitable iterative method (appendix A).
Evidently, the histograms of the values 
calculated in the course of a numerical
simulation approximate the gap 
distribution up to statistical errors.
In this section, we report the results of such numerical studies and show that
the data are well described by a few simple empirical laws.

\subsection 3.1 Simulation parameters

The simulations listed in table~1 are part of an ongoing study
of two-flavour QCD in the chiral regime
[\ref{NextPaper}].
Technically the project is based on the use of the Schwarz-preconditioned
HMC simulation algorithm introduced in ref.~[\ref{SchwarzAlgorithmII}], 
which allows the theory to be simulated in a range of lattices and 
quark masses that was, in practice, inaccessible so far. 

Except for the last run in table~1,
the unimproved Wilson theory was simulated,
at inverse bare coupling $\beta$ and hopping parameter
$\kappa=(8+2am_0)^{-1}$. In run $D_1$ the coefficient $\csw$ of the 
Sheikholeslami--Wohlert improvement term [\ref{SW},\ref{OaImp}] was set
to the value determined by the ALPHA collaboration [\ref{JansenSommer}].
A number $\Ncfg$ of statistically decorrelated gauge-field configurations
was generated in each case and later used for the
calculation of the gap distributions.

\midinsert
\newdimen\digitwidth
\setbox0=\hbox{\rm 0}
\digitwidth=\wd0
\catcode`@=\active
\def@{\kern\digitwidth}
\tablecaption{Simulation runs included in this study} 
\vskip-1.0ex
$$\vbox{\settabs\+&%
                  xxxxxxx&xx&
                  xxxxxxxxx&xx&
                  xxxxx&xx&
                  xxxxxxxx&xx&
                  xxxxxxxx&xx&
                  xxxxxx&xx&
                  xxxxxxxxxxxx&\cr
\thicktablerule
\vskip1.0ex
                \+& \hfill Run\hfill
                 && \hfill Lattice\hfill
                 && \hfill $\beta$\hfill
                 && \hfill $\csw$\hfill
                 && \hfill $\kappa$\hfill
                 && \hfill $\Ncfg$\hfill
                 && \hfill $a\mpi$\hfill
                 &\cr
\vskip1.0ex
\thintablerule
\vskip1.5ex
  \+& \hfill $A_1$\hfill
  &&  \hfill $32\times24^3$\hfill
  &&  \hfill $5.6$\hfill 
  &&  \hfill $0$\hfill
  &&  \hfill $0.15750$\hfill
  &&  \hfill $@64$\hfill
  &&  \hfill $0.2744(21)$\hfill
  &\cr
\vskip0.3ex
  \+& \hfill $A_2$\hfill
  &&  \hfill $$\hfill
  &&  \hfill $$\hfill 
  &&  \hfill $$\hfill
  &&  \hfill $0.15800$\hfill
  &&  \hfill $109$\hfill
  &&  \hfill $0.1969(16)$\hfill
  &\cr
\vskip0.3ex
  \+& \hfill $A_3$\hfill
  &&  \hfill $$\hfill
  &&  \hfill $$\hfill 
  &&  \hfill $$\hfill
  &&  \hfill $0.15825$\hfill
  &&  \hfill $100$\hfill
  &&  \hfill $0.1554(31)$\hfill
  &\cr
\vskip0.3ex
  \+& \hfill $A_4$\hfill
  &&  \hfill $$\hfill
  &&  \hfill $$\hfill 
  &&  \hfill $$\hfill
  &&  \hfill $0.15835$\hfill
  &&  \hfill $100$\hfill
  &&  \hfill $0.1204(44)$\hfill
  &\cr
\vskip0.3ex
  \+& \hfill $B_1$\hfill
  &&  \hfill $64\times32^3$\hfill
  &&  \hfill $5.8$\hfill 
  &&  \hfill $0$\hfill
  &&  \hfill $0.15410$\hfill
  &&  \hfill $100$\hfill
  &&  \hfill $0.1965(8)@$\hfill
  &\cr
\vskip0.3ex
  \+& \hfill $B_2$\hfill
  &&  \hfill $$\hfill
  &&  \hfill $$\hfill 
  &&  \hfill $$\hfill
  &&  \hfill $0.15440$\hfill
  &&  \hfill $101$\hfill
  &&  \hfill $0.1481(11)$\hfill
  &\cr
\vskip0.3ex
  \+& \hfill $C_1$\hfill
  &&  \hfill $64\times24^3$\hfill
  &&  \hfill $5.6$\hfill 
  &&  \hfill $0$\hfill
  &&  \hfill $0.15800$\hfill
  &&  \hfill $116$\hfill
  &&  \hfill $0.1986(10)$\hfill
  &\cr
\vskip0.3ex
  \+& \hfill $D_1$\hfill
  &&  \hfill $48\times24^3$\hfill
  &&  \hfill $5.3$\hfill 
  &&  \hfill $1.90952$\hfill
  &&  \hfill $0.13550$\hfill
  &&  \hfill $104$\hfill
  &&  \hfill $0.3265(11)$\hfill
  &\cr
\vskip1.5ex
\thicktablerule
}
$$
\vskip-2ex
\endinsert

\vfill\eject

Details of the simulations, the computation of the quark masses
and other physical quantities will be given in a forthcoming
publication [\ref{NextPaper}]. 
Based on calculations of 
the Sommer reference scale $r_0$ [\ref{SommerRadius}], the lattice
spacing in the Wilson theory is estimated to be about $0.08$ and $0.06$ fm
at $\beta=5.6$ and $5.8$, respectively, while
it is roughly $0.09$ fm in the improved theory at $\beta=5.3$.
In physical units the spatial sizes of the lattices 
in table~1 are thus close to $2$ fm in all cases. 

As usual the conversion from lattice to physical units
is ambiguous, but in this paper the latter
serve for the purpose of illustration only.
The pion masses quoted in the last column
of table~1, for example, cover a range from $676$ to about $294$ MeV
in the case of the runs $A_1$--$A_4$.

\topinsert
\vbox{
\vskip0.0cm
\epsfxsize=6.0cm\hskip3.0cm\epsfbox{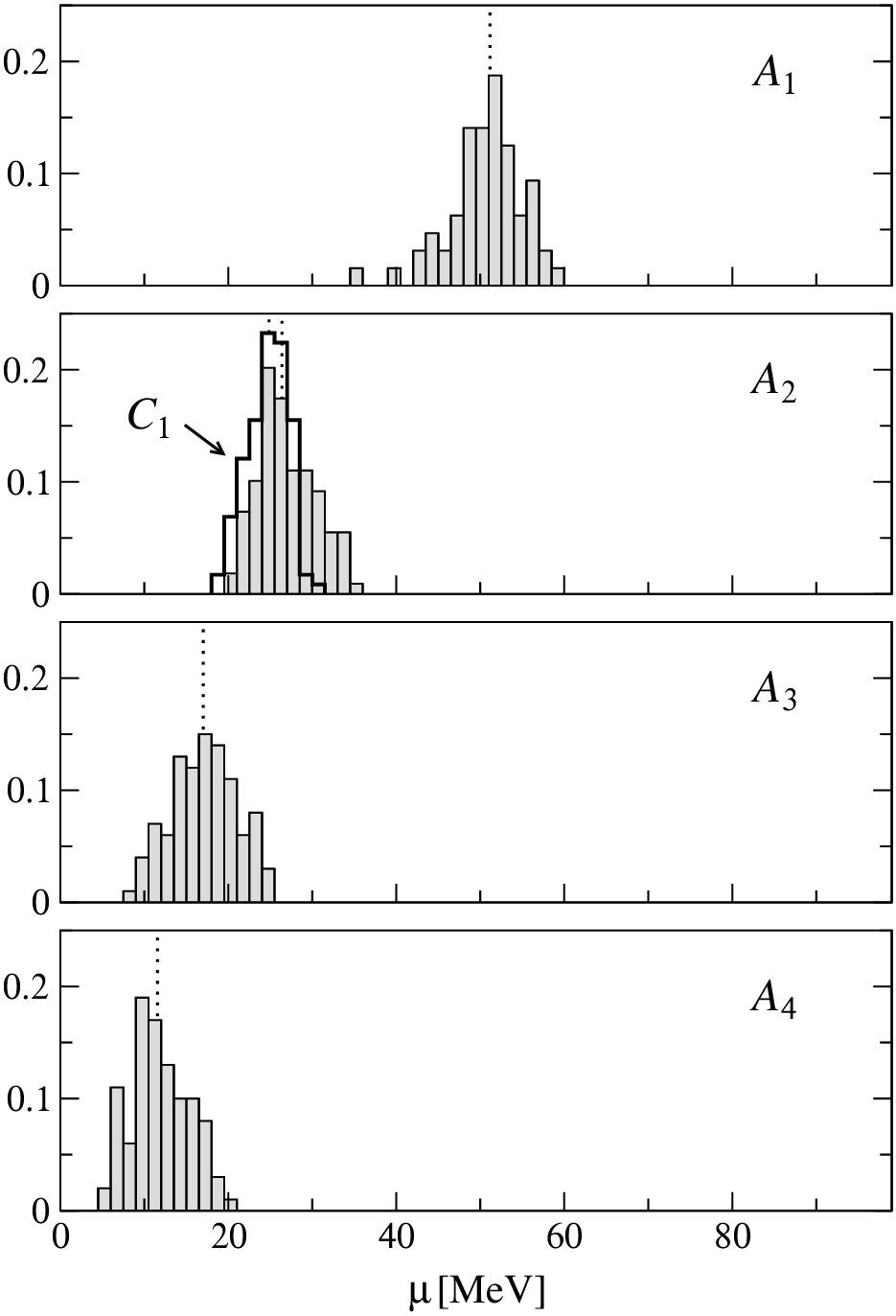}
\vskip0.3cm
\figurecaption{%
Normalized histograms of the spectral gap $\gap$,
as obtained in the runs $A_1$--$A_4$ and $C_1$.
The bin size is $1.5$ MeV and 
the dotted vertical lines indicate
the position of the median $\gapbar$ of the distributions.
}
\vskip0.0cm
}
\endinsert

\subsection 3.2 Mass-dependence of the gap distribution

The calculated distributions of the gap in the Wilson 
theory at $\beta=5.6$ are plotted in fig.~2. A characteristic
feature of these distributions is that their shape does not show
a strong dependence on the quark mass.
Basically they are shifted
to smaller values when the quark mass decreases. 
All these distributions are clearly separated from the origin,
although the smallest mass may be quite close to where
the unstable regime begins.

\topinsert
\vbox{
\vskip0.0cm
\epsfxsize=7.0cm\hskip2.0cm\epsfbox{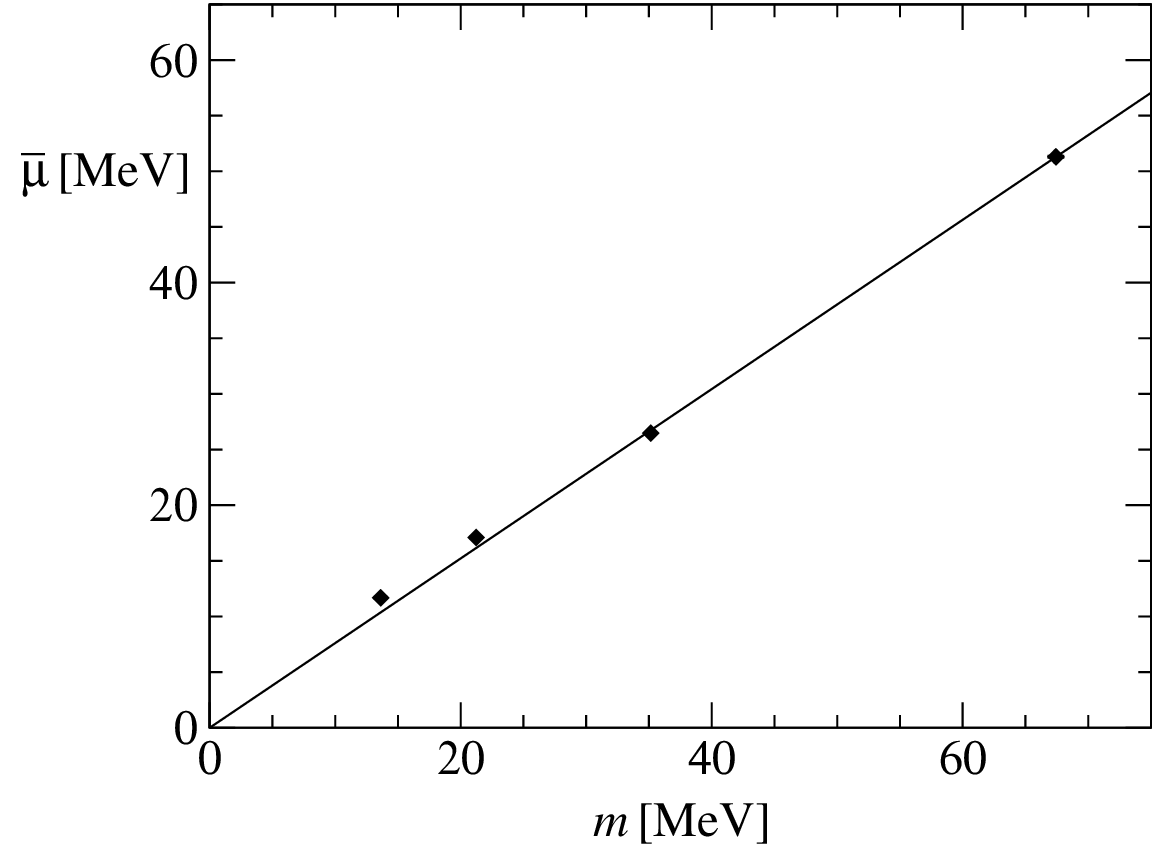}
\vskip0.4cm
\figurecaption{%
Median $\gapbar$ of the gap distribution 
on the $32\times24^3$ lattice (runs $A_1$--$A_4$)
as a function of the bare current-quark mass $m$.
The straight line from the point
at the largest mass to the origin is drawn to guide the eye.
Statistical errors are negligibly small on the scale of this plot.
}
\vskip0.0cm
}
\endinsert

The distributions are roughly symmetric about their median (dotted
lines in fig.~2), which is practically also the point where they
assume their maximal value. As shown in fig.~3,
the median scales approximately
linearly with the current-quark mass. A small but significant
effect is seen at the lighter quark masses, where the median
is pushed to slightly larger values
with respect to the straight scaling curve.

\topinsert
\vbox{
\vskip0.0cm
\epsfxsize=7.0cm\hskip2.0cm\epsfbox{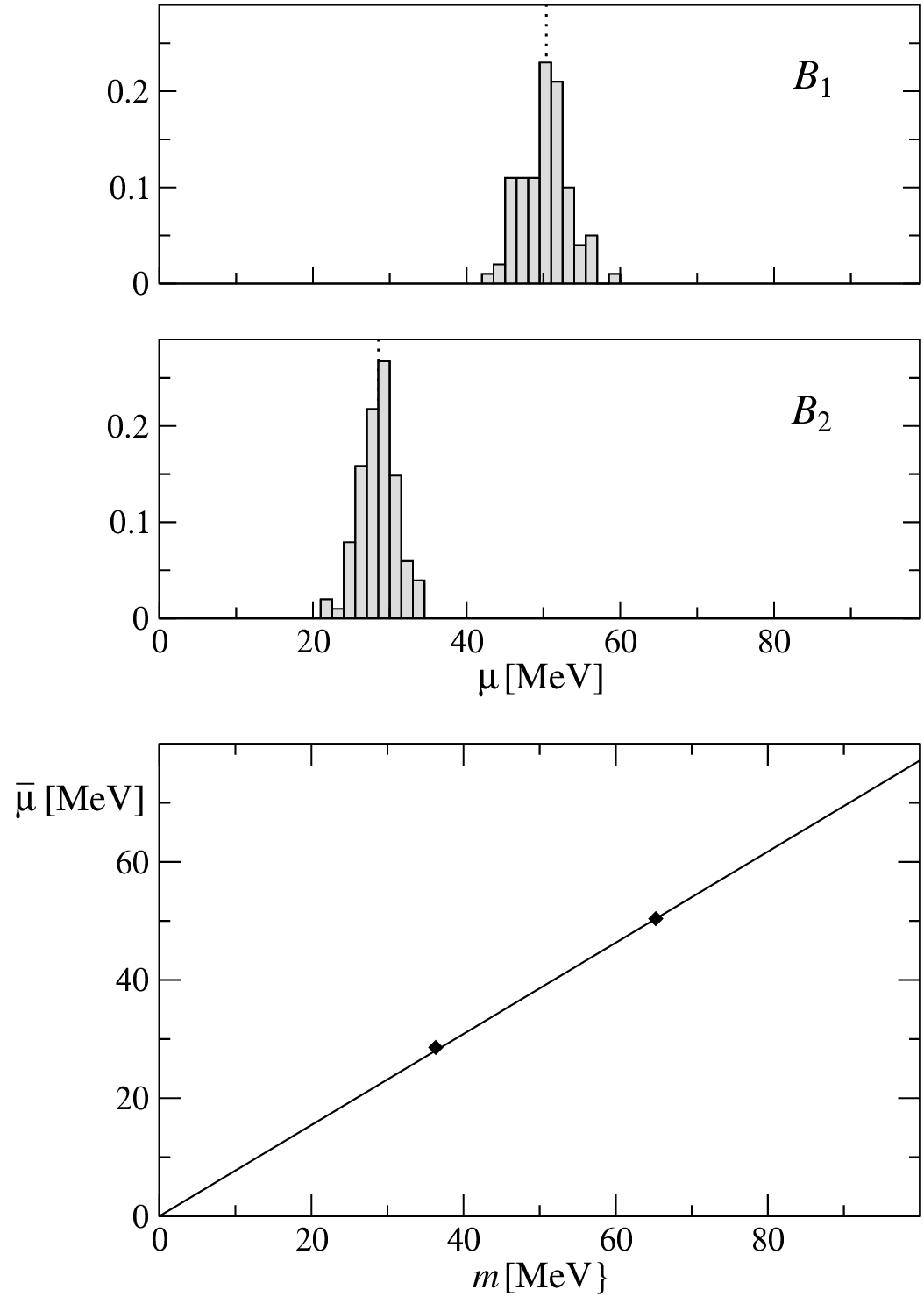}
\vskip0.4cm
\figurecaption{%
Gap distributions and mass dependence of the median $\gapbar$
on the $64\times32^3$ lattices (runs $B_1$ and $B_2$).
The bin size is the same as in fig.~2.
}
\vskip0.0cm
}
\endinsert

On the big lattices at $\beta=5.8$, the situation is essentially the same,
although here only two simulations have been completed so far
(see fig.~4).
In particular, the quark masses in these two runs are in a 
range where a nearly perfect scaling of 
the median as a function of the quark mass is again observed.

\subsection 3.3 Statistical fluctuations of the gap

An important qualitative result of 
our empirical studies is that the fluctuations
of the gap become smaller when the four-dimensional volume $V$
of the lattice increases.
The simulation $C_1$, for example,
was performed at exactly the same coupling
and quark mass as the simulation $A_2$, but on a larger lattice. 
As can be seen from fig.~2, the width of the gap distribution
obtained in run $C_1$ is visibly reduced with respect to the one
obtained in run $A_2$.

A simple heuristic argumentation that may explain this effect
goes as follows. Let $U$ be a given lattice gauge field, 
$\mu$ the lowest eigenvalue of $\left|\hDm\right|$ and $\psi$
the associated eigenvector normalized to unity. Consider a
small random fluctuation $U+\delta U$ of the gauge field. 
To first order, the
change $\delta\gap$ in the gap is then given by
\equation{
  \delta\gap=a^4\sum_x\psi(x)^{\dagger}\left[\delta\hDm\psi\right](x).
  \enum
}
At each point $x$, there are 
eight contributions to this sum, which are
proportional to the 
random variation $\delta U$ of the gauge field on the links attached there.
Now if $\psi$ extends over the whole lattice, i.e.~if $\psi$
it is not a localized mode, the size of these terms is 
proportional to $a^3/V$. On average this implies
$\langle\delta\gap\rangle=0$ and 
$\langle(\delta\gap)^2\rangle\propto a^2/V$.

While this kind of reasoning is extremely superficial,
the width $\sigma$ of the numerically determined 
gap distributions appears to scale 
in the suggested way (see fig.~5)%
$\,$\footnote{$\dagger$}{\footnotefont%
The standard deviation of $\gap$ is subject to potentially
large statistical uncertainties, because the tails of the 
gap distributions are poorly sampled.
We therefore define the width of the distributions through
$\sigma=\frac{1}{2}(v-u)$,
where $[u,v]$ is the smallest range in $\gap$, which contains  
more than $68.3\%$ of the data.}.
An interesting special case is that of the distributions obtained in the 
runs $A_2$ and $C_1$, where, 
as already mentioned, all parameters were the same
except for the lattice volumes. The scaled widths $\sigma\sqrt{V}/a$, 
however,
agree very well with each other.
The approximate scaling law
\equation{
  \sigma\simeq{a\over\sqrt{V}}
  \enum
}
is actually consistent with all the data shown in fig.~5.

\topinsert
\vbox{
\vskip0.0cm
\epsfxsize=6.5cm\hskip2.5cm\epsfbox{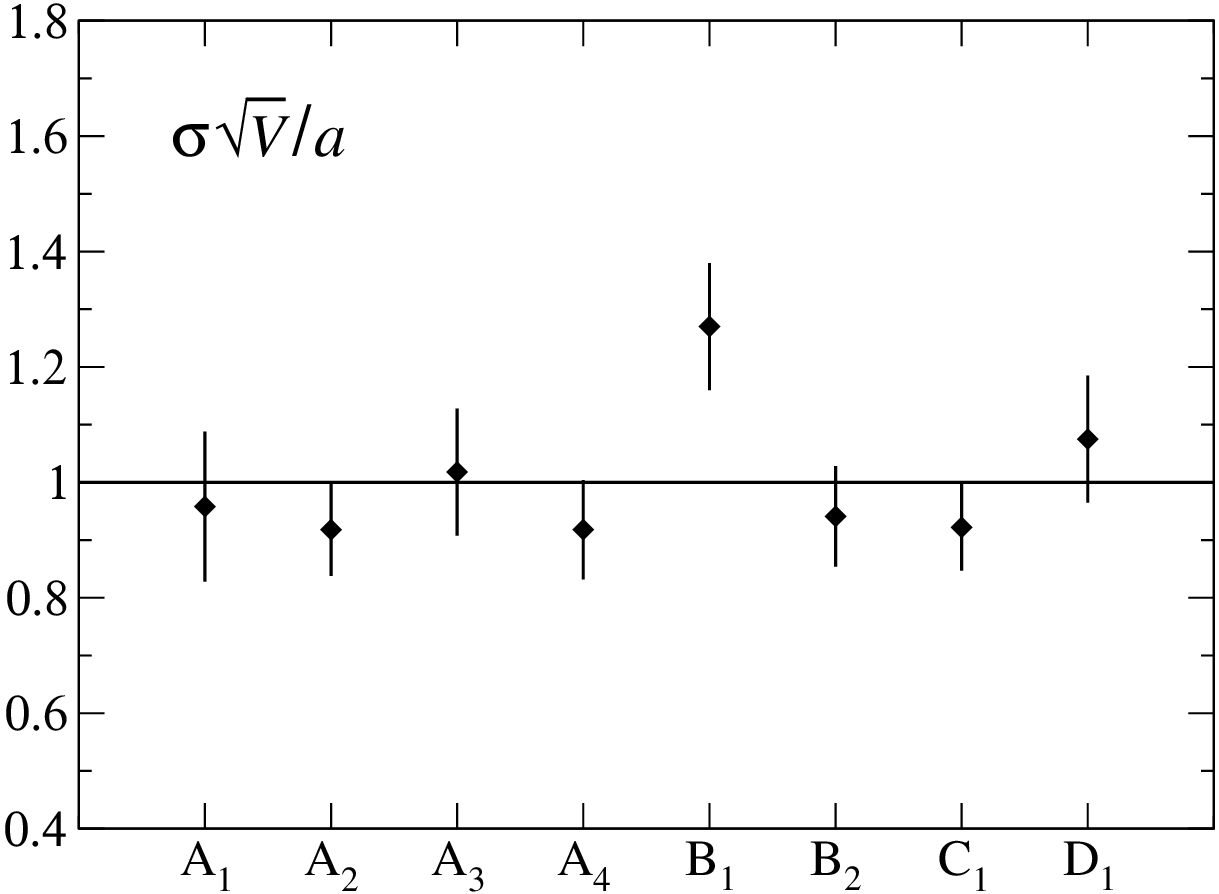}
\vskip0.4cm
\figurecaption{%
Width $\sigma$ of the gap distributions, given in units of $a/\sqrt{V}$,
as obtained in runs $A_1$--$D_1$. Note that the combination $\sigma\sqrt{V}/a$
is dimensionless and can be computed directly from the lattice data,
without intermediate conversion to physical units.
The statistical errors were determined
using the bootstrap method.
}
\vskip0.0cm
}
\endinsert

Somewhat surprisingly, the spectral gap thus seems to behave thermodynamically,
like a free-energy density for example, with probability distribution
\equation{
   p(\gap)\propto\exp
   \left\{-{V\over2a^2}\left(\gap-\langle\gap\rangle\right)^2\right\}.
   \enum
}
The fact that $\sigma$ is proportional to the lattice spacing
suggests, on the other hand, that the observed widths result
from short-distance fluctuations of the gauge field. These 
fluctuations scale more slowly with the volume
than those expected in lattice
theories that preserve chiral symmetry, where the eigenvalue distributions
are universally
computable, using chiral perturbation theory or random matrix theory.

\section 4. Spectral gap in infinite volume

Analytical calculations of the gap distribution would evidently
be very welcome at this point, but since chiral symmetry
is violated in the Wilson theory, it is unclear how 
such calculations would proceed.
The spectral density of the Dirac operator 
in infinite volume is somewhat more accessible
and provides a useful reference for the 
situation on the lattices that can be simulated.

\subsection 4.1 Spectral density

Let $\alpha_1,\alpha_2,\ldots$ be the eigenvalues of $\hDm^2$, ordered
in ascending order and counting multiplicities. 
In the following we will be interested in the spectral density
\equation{
  \rho(\alpha)=\lim_{V\to\infty}{1\over V}\sum_{k\geq1}
  \left\langle\delta(\alpha-\alpha_k)\right\rangle
  \enum
}
in infinite volume. It is, incidentally, possible to prove
rigorously that the thermodynamic limit (4.1) exists,
using a general argument based on a decomposition
of the lattice into large blocks [\ref{vanHemmen}].

In the continuum theory, or if chiral symmetry is preserved 
by the lattice theory,
the spectral density vanishes
at $\alpha<m^2$. Moreover, for slightly larger values of $\alpha$,
the Banks--Casher relation [\ref{BanksCasher}] implies the 
asymptotic form
\equation{
  \rho(\alpha)\mathrel{\mathop=_{\alpha>m^2}}
  {\Sigma\over\pi\sqrt{\alpha-m^2}}+\rmO(1),
  \enum
}
where $\Sigma$ denotes the quark condensate.
The spectral density thus has a well-defined threshold 
at $\alpha=m^2$ in these cases (see fig.~6).

In the following a {\it working hypothesis}\/ is that 
there is a similar threshold $\thresh>0$ 
in the Wilson theory or its O($a$)-improved version (whichever
is considered). 
We shall not need to know the shape
of the spectral density in the vicinity of the threshold, but it must
be guaranteed that $\rho(\alpha)$ vanishes if $\alpha<\thresh$
and that the density is non-zero at or immediately above $\thresh$. 
It is important to understand
that $\thresh$ is the point where the dense spectrum
begins, while it is perfectly possible that the Wilson--Dirac operator in 
finite volume has eigenvalues significantly smaller than $\sqrt{\thresh}$. 
Any part of the spectrum with less than O($V$) modes per energy bin 
is in fact suppressed by the factor $1/V$ in the limit (4.1).

\topinsert
\vbox{
\vskip0.0cm
\epsfxsize=6.5cm\hskip2.5cm\epsfbox{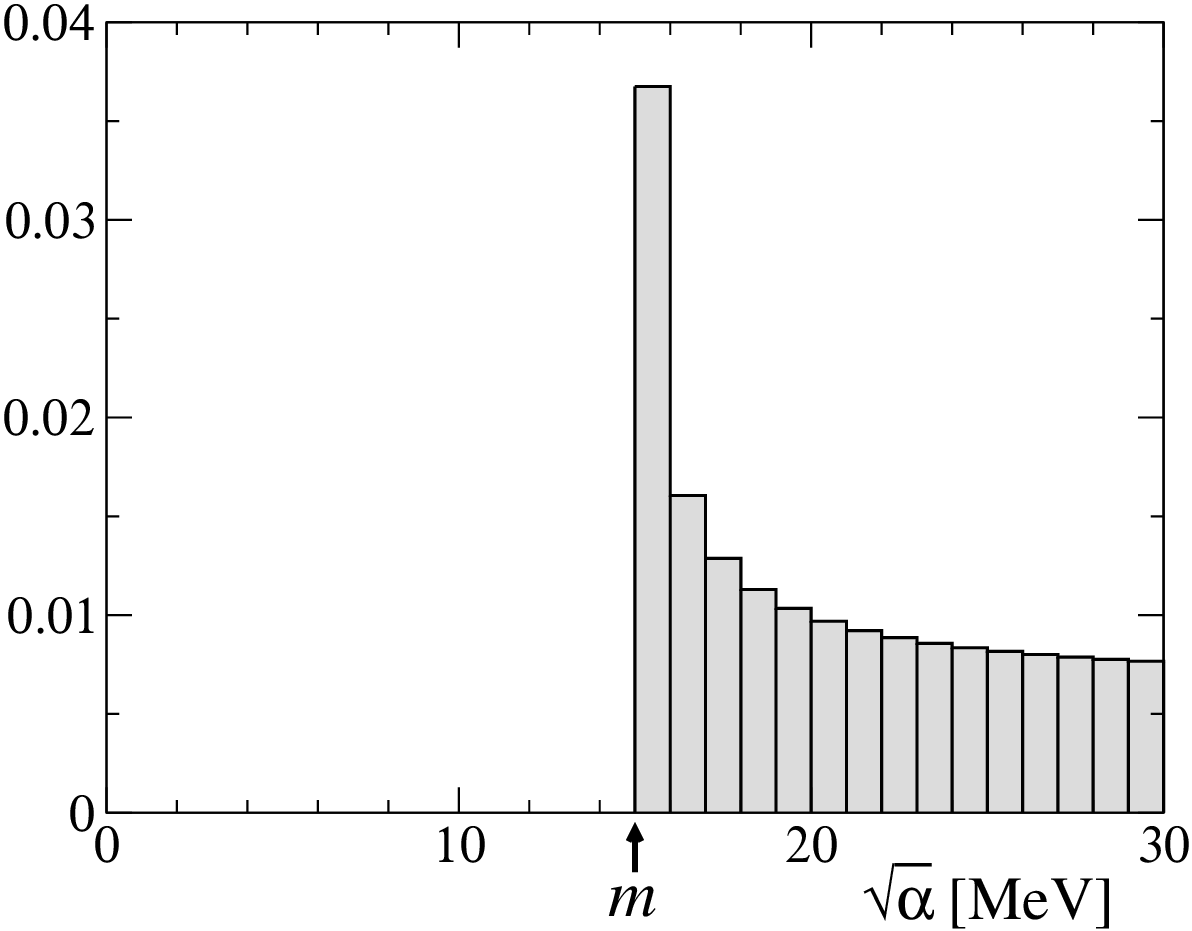}
\vskip0.4cm
\figurecaption{%
Number of eigenvalues per bin and $\fm^4$ at the low end
of the spectral density in the continuum theory
[cf.~eq.~(4.2)].
In this plot, $m=15$ MeV and $\Sigma=(250\,\MeV)^3$ was assumed.
}
\vskip0.0cm
}
\endinsert

\subsection 4.2 Adding valence quarks

To be able to relate the spectral density to correlation functions
of local fields, we will need to consider partially quenched
QCD, where $2N$ valence quarks are added to the theory
[\ref{BernardGolterman},\ref{SharpeShoresh}].
The fermion action then becomes
\equation{
  S_{\rm F}=a^4\sum_x\left\{\sum_{r=1}^{2N+2}\psibar_{r}(x)D_m\psi_r(x)+
  \sum_{k=1}^N\left|D_m\phi_k(x)\right|^2\right\},
  \enum
}
where $\psibar_r,\psi_r$ are the quark fields 
($2$ sea quarks plus $2N$ valence quarks) and $\phi_k$
the pseudo-fermion fields that are required to cancel 
the valence-quark determinants. This defines a well-behaved
euclidean field theory, whose renormalization can be 
expected to follow the usual rules. In particular,
the obvious
$\rmU(2N+2)\times\rmU(N)$ flavour symmetry
and some less obvious graded symmetries
restrict the possible counterterms to the 
naively expected ones.

We now also introduce the bare scalar and pseudo-scalar densities
\equation{
  S_{rs}=\psibar_r\psi_s,\qquad P_{rs}=\psibar_r\dirac{5}\psi_s.
  \enum
}
These fields renormalize in the standard manner, i.e.~for unequal flavours
$r,s$, the renormalized operators are $\ZS S_{rs}$ and $\ZP P_{rs}$,
where the renormalization constants 
$\ZS$ and $\ZP$ are flavour-independent and can be
taken to be the same as the ones
in the theory without valence quarks.

\subsection 4.3 Resolvent \& moments of the spectral density

The resolvent 
\equation{
  R(z)=\int_{\thresh}^{\infty}
  \rmd\alpha\,{\rho(\alpha)\over\alpha^2(z-\alpha)},
  \enum
}
is an analytic function of $z$ with a cut along the real axis
from $z=\thresh$ to some value proportional to $1/a^2$.
For later convenience, a factor $\alpha^{-2}$ 
is included in the integral to improve its convergence 
at large eigenvalues.
Evidently, the spectral density is proportional to the discontinuity
across the cut, and so can be uniquely recovered from the resolvent
if the latter is known [\ref{OsbornPQChPT},\ref{DamgaardPQChPT}].

For $\left|z\right|<\thresh$, the resolvent may be expanded in 
a convergent power series,
\equation{
  R(z)=\sum_{k=0}^{\infty}M_kz^k,
  \qquad
  M_k=-\int_{\thresh}^{\infty}
  \rmd\alpha\,{\rho(\alpha)\over\alpha^{k+3}},
  \enum
}
with coefficients $M_k$ that can be written as
\equation{
  M_k=a^{4n-4}\sum_{x_1,\ldots,x_{n-1}}
  \left\langle
  P_{12}(x_1)P_{23}(x_2)\ldots P_{n1}(0)
  \right\rangle,
  \qquad n=2k+6.
  \enum
}
This formula assumes that there are at least $n$ quarks, but
since each moment $M_k$ may be considered separately, 
a sufficient number of valence quarks can always be added
to the theory.

\subsection 4.4 Renormalization 

Equations (4.5)--(4.7) connect the spectral density
to the basic field-theoretic correlation functions whose renormalization 
properties are well understood.
The suggestion is then that 
the moments $M_k$ get renormalized through multiplication by the 
factor $(\ZP)^n$.
Since the sum over the coordinates $x_1,\ldots,x_{n-1}$ 
in eq.~(4.7) includes
the short-distance regions, it is, however, not totally obvious
that the moments thus renormalized will indeed have a well-defined 
continuum limit. 

In the continuum theory, and when inserted in correlation functions,
the product $P_{12}(x_1)P_{23}(x_2)$, for example, has a
short-distance expansion 
\equation{
   P_{12}(x_1)P_{23}(x_2)\mathrel{\mathop\sim_{x_1\to x_2}}
   C(x_1-x_2)S_{13}(x_2)+\ldots,
   \enum
}
where the Wilson coefficient $C(x)$ diverges like 
$\left|x\right|^{-3}$ (up to logarithms).
This singularity is integrable and does not give rise to any
additional ultraviolet divergencies. 
As it turns out, all
short-distance singularities of the correlation functions in eq.~(4.7)
are in fact integrable$\,$\footnote{$\dagger$}{\footnotefont%
When all coordinates $x_1,\ldots,x_{n-1}$ are scaled to zero,
the operator product reduces to a coefficient function times the unit 
operator. The degree of divergence of the integral is $4-n$ in this case,
and convergence is thus guaranteed since $n\geq6$.}.
The ultraviolet divergencies of the moments are thus completely
cancelled by the renormalization factor $(\ZP)^n$ and by the
usual renormalization of the gauge coupling and the quark mass.

Recalling the expansion (4.6) of the resolvent $R(z)$ and its relation
to the spectral density, it now follows that
\equation{
  \rhoR(\alpha)=\ZP^2\rho(\ZP^2\alpha)
  \enum
}
has a universal continuum limit (once a particular renormalization condition
for the pseudo-scalar density is adopted). 
The same must then also apply to the re\-nor\-ma\-lized threshold
\equation{
  \threshR=\ZP^{-2}\thresh,
  \enum
}
since the threshold is a property of the spectral density.

Without proof we mention in passing that
the improved renormalized density in the 
O($a$)-improved theory is again given by eq.~(4.9),
provided $\ZP$ is replaced by
\equation{
  \ZP{1+\bp a\mq\over 1+\bpp a\mq},
  \enum
}
where $\bp$ and $\bpp$ are improvement coefficients
and $\mq$ the additively renormalized bare quark mass
(the notation is as in ref.~[\ref{OaImp}]). 
The correction proportional to $\bpp$ in this expression 
is in fact all that is needed to cancel the terms of order $a$ that 
arise from the integrations over the 
short-distance singularities of the correlation function 
on the right of eq.~(4.7).

\subsection 4.5 Relation to the current-quark mass

The bare current-quark mass $m$ is usually determined through
vacuum-to-pion matrix elements of the
isovector axial current and density. From the renormalization 
properties of the matrix elements it then follows that 
the renormalized quark mass is given by
\equation{
  \mr=\ZA\ZP^{-1}m,
  \enum
}
where $\ZA$ denotes the axial-current renormalization constant.

Once a definite renormalization condition for the isovector
axial density is adopted, the threshold $\threshR$ and the quark mass
$\mr$ become physical quantities.
In particular, in the continuum limit we have
\equation{
  {\sqrt{\threshR}\over\mr}=1,
  \enum
}
not only when the limit is reached from a lattice theory 
that preserves chiral symmetry, because
this ratio is dimensionless, unambiguously normalized and therefore
independent of the regularization. In terms of the bare quantities
this implies $\sqrt{\thresh}=\ZA m$, up to corrections of O($a$)
and independently of the normalization convention for the renormalized 
axial density 
[the renormalization constant $\ZP$ cancels in the ratio (4.13)].

As already mentioned, the spectral gap in finite volume may not
be related in any simple way to the threshold of the spectral density in 
the thermodynamic limit. It is nevertheless instructive to 
compare the median $\gapbar$ of the gap distributions discussed
in sect.~3 with the threshold $\ZA m$.

\midinsert
\newdimen\digitwidth
\setbox0=\hbox{\rm 0}
\digitwidth=\wd0
\catcode`@=\active
\def@{\kern\digitwidth}
\tablecaption{Comparison of the median $\gapbar$ with $\ZA m$} 
\vskip-1.0ex
$$\vbox{\settabs\+&%
                  xxxxxxx&xx&
                  xxxxxxxxx&xxx&
                  xxxxxxxxxx&xx&
                  xxxxxxxxxxxxxxxx&xx&
                  xxxxxxxxxxx&\cr
\thicktablerule
\vskip1.0ex
                \+&  \hfill Run\hfill
                  && \hfill$\ZA$\hfill 
                  && \hfill\hskip-1ex$\gapbar/m$\hfill
                  && \hfill$\gapbar-\ZA m$ [MeV]\hfill
                  && \hfill$\left\langle\Delta\right\rangle$ [MeV]\hfill
                  &\cr
\vskip1.0ex
\thintablerule
\vskip1.5ex
  \+& \hfill $A_1$\hfill
  &&  \hfill $0.77(2)^a$\hfill
  &&  \hfill $0.76(1)$\hfill 
  &&  \hfill $-0.5(5)$\hfill
  &&  \hfill $2.8(1)@$\hfill
  &\cr
\vskip0.3ex
  \+& \hfill $A_2$\hfill
  &&  \hfill $$\hfill
  &&  \hfill $0.75(1)$\hfill 
  &&  \hfill $-0.6(4)$\hfill
  &&  \hfill $3.2(1)@$\hfill
  &\cr
\vskip0.3ex
  \+& \hfill $A_3$\hfill
  &&  \hfill $$\hfill
  &&  \hfill $0.80(3)$\hfill 
  &&  \hfill $\phantom{-}0.6(5)$\hfill
  &&  \hfill $3.5(1)@$\hfill
  &\cr
\vskip0.3ex
  \+& \hfill $A_4$\hfill
  &&  \hfill $$\hfill
  &&  \hfill $0.85(3)$\hfill 
  &&  \hfill $\phantom{-}1.1(4)$\hfill
  &&  \hfill $3.9(1)@$\hfill
  &\cr
\vskip0.3ex
  \+& \hfill $B_1$\hfill
  &&  \hfill $0.78(2)^a$\hfill
  &&  \hfill $0.77(1)$\hfill 
  &&  \hfill $-0.3(3)$\hfill
  &&  \hfill $1.91(6)$\hfill
  &\cr
\vskip0.3ex
  \+& \hfill $B_2$\hfill
  &&  \hfill $$\hfill
  &&  \hfill $0.79(1)$\hfill 
  &&  \hfill $-0.2(3)$\hfill
  &&  \hfill $2.19(6)$\hfill
  &\cr
\vskip0.3ex
  \+& \hfill $C_1$\hfill
  &&  \hfill $0.77(2)^a$\hfill
  &&  \hfill $0.72(1)$\hfill 
  &&  \hfill $-1.7(3)$\hfill
  &&  \hfill $2.18(7)$\hfill
  &\cr
\vskip0.3ex
  \+& \hfill $D_1$\hfill
  &&  \hfill $0.75(1)^b$\hfill
  &&  \hfill $0.68(1)$\hfill 
  &&  \hfill $-5.4(5)$\hfill
  &&  \hfill $2.2(1)@$\hfill
  &\cr
\vskip1.5ex
\thicktablerule
\vskip1.8ex
\hbox{\hskip0.1em\footnotefont$^a$ 
RI--MOM method [\ref{BecirevicEtAl}]}
\hbox{\hskip0.1em\footnotefont$^b$ 
Schr\"odinger functional chiral Ward identity [\ref{DellaMorteEtAl}]}
}
$$
\vskip-2ex
\endinsert
\vfill\eject

The numerically computed values of $\gapbar/m$ quoted in table~2
actually agree quite well with the available estimates of $\ZA$.
There are, however, significant differences in the last two rows
of the table (runs $C_1$ and $D_1$), 
which underlines the fact that there is currently no 
solid theoretical understanding of 
the gap distributions in finite volume.
On the other hand, the absolute deviation of the median
from the threshold is, in most cases, smaller
than the average splitting 
$\left\langle\Delta\right\rangle$
of the first four eigenvalues of  $\left|\hDm\right|$ 
(see table~2; the figures quoted in the 
fourth column do not include the error on $\ZA$).
In particular, the data are consistent
with the working hypothesis on which our 
argumentation relied (cf.~subsect.~4.1).

\section 5. Conclusions

As explained in sect.~2,
numerical simulations of the Wilson theory can be expected to 
be stable if the distribution of the spectral gap 
of the lattice Dirac operator is well separated from the origin.
The range of stability may be defined through 
the inequality $\gapbar\geq3\sigma$, for example, where, as before, 
$\gapbar$ and $\sigma$ denote the median and width of the distribution.
Using the empirical
relations $\gapbar\simeq Z m$ and $\sigma\simeq a/\sqrt{V}$,
this bound becomes $m\geq3a/Z\sqrt{V}$, which shows that 
the accessible range of quark masses depends on both the 
lattice spacing and the lattice size.

Another form of the stability bound is obtained by 
multiplication with the ratio $B=\mpi^2/2m$, which is 
known to be practically independent of the quark mass
[\ref{SchwarzReview},\ref{NextPaper}].
For lattices of size $2L\times L^3$, this leads to the inequality
\equation{
  \mpi L\geq\sqrt{3\sqrt{2}aB/Z}.
  \enum
}
In particular, from our numerical studies
of the (unimproved) Wilson theory, we deduce the stability bounds
\equation{
  \mpi L\geq\cases{$2.8$ & at $a=0.08$ fm,\cr
                   \noalign{\vskip1ex}
                   $2.3$ & at $a=0.06$ fm,\cr}
  \enum
}
where the mass dependence of $B$ and $Z$ was neglected for simplicity.
Note that the pion mass 
in these formulae is the one in infinite volume,
and not the possibly much larger mass
computed on a lattice of size $L$.

It follows from these results that
the range of stability includes all lattices 
where, say, $a\leq0.1$ fm, $L\geq2$ fm and $\mpi L\geq3$.
Simulations of the Wilson theory on such lattices,
using the known simulation algorithms,
are thus expected to be safe from the problems mentioned in sect.~2.
The O($a$)-improved theory
is likely to behave in the same way, but so far this has only 
been checked on a single lattice with spacing 
$a=0.09$ fm. If the median and width of the gap distribution are 
assumed to scale as in the unimproved theory, the stability
bound deduced from the simulation of this lattice 
is $\mpi L\geq 3.2$. Extensive
studies of the improved theory will evidently be required to 
confirm this result, which is very much in line with the bounds (5.2).

\vskip1.5ex
The numerical simulations reported in this paper 
were performed on PC clusters at
CERN, the Centro Enrico Fermi and 
the Institut f\"ur Theoretische Physik der Universit\"at Bern
(with a contribution from the Schweizerischer Nationalfonds).
We are grateful to all these institutions for the continuous support
given to this project.
L.~G.~acknowledges partial support from the EU under contract 
HPRN-CT-2002-00311 (EURIDICE).

\appendix A. Chebyshev accelerated subspace iteration

On a finite lattice, the few lowest eigenvalues of $A=\hDm^2$ 
can be computed numerically by minimizing the associated Ritz functional, 
for example [\ref{EvaI},\ref{EvaII}]. 
This method is relatively tolerant of
rounding errors, which is an important advantage on 
computers that do not support double-precision arithmetic.
Otherwise there is a choice of algorithms
that can be significantly faster. Subspace iteration with
Chebyshev acceleration and eigenvector locking is one of them,
and it is our aim, in the following paragraphs, to describe
this method in some detail
(see ref.~[\ref{SubspaceI}], for example).

\subsection A.1 Power method

A tight lower bound on the largest eigenvalue of $A$ can be obtained
by repeatedly applying the operator to a random quark field. 
Starting from some arbitrary field $\psi$ with unit norm,
the recursion 
\equation{
   \chi=A\psi,
   \qquad
   \psi=\chi/\|\chi\|,
   \enum
} 
systematically enhances the upper spectral components of the field.
The norm $\|\chi\|$ then provides
an increasingly accurate estimate of the largest eigenvalue.
In practice $20$ iterations or so are usually sufficient for an accuracy 
better than
$5\%$.

When $A$ is replaced by the shifted operator $c-A$, where 
$c>\frac{1}{2}\|A\|$ is some fixed number, the power iteration
converges to $c$ minus the lowest eigenvalue of $A$.
The latter can thus be calculated in this way.
However, since the low eigenvalues of $A$ are typically orders of 
magnitude smaller than $c$, accurate results
are obtained only after a very large number of iterations.

The power method is therefore not recommended for the computation of
the low eigenvalues of $A$,
unless it is combined with an acceleration technique. In particular,
the shifted operator can be replaced by a Chebyshev polynomial and
the iteration may be extended to a subspace of quark fields.
Both of these modifications 
lead to significantly improved convergence rates.

\topinsert
\vbox{
\vskip0.0cm
\epsfxsize=8.0cm\leftline{\hfill\epsfbox{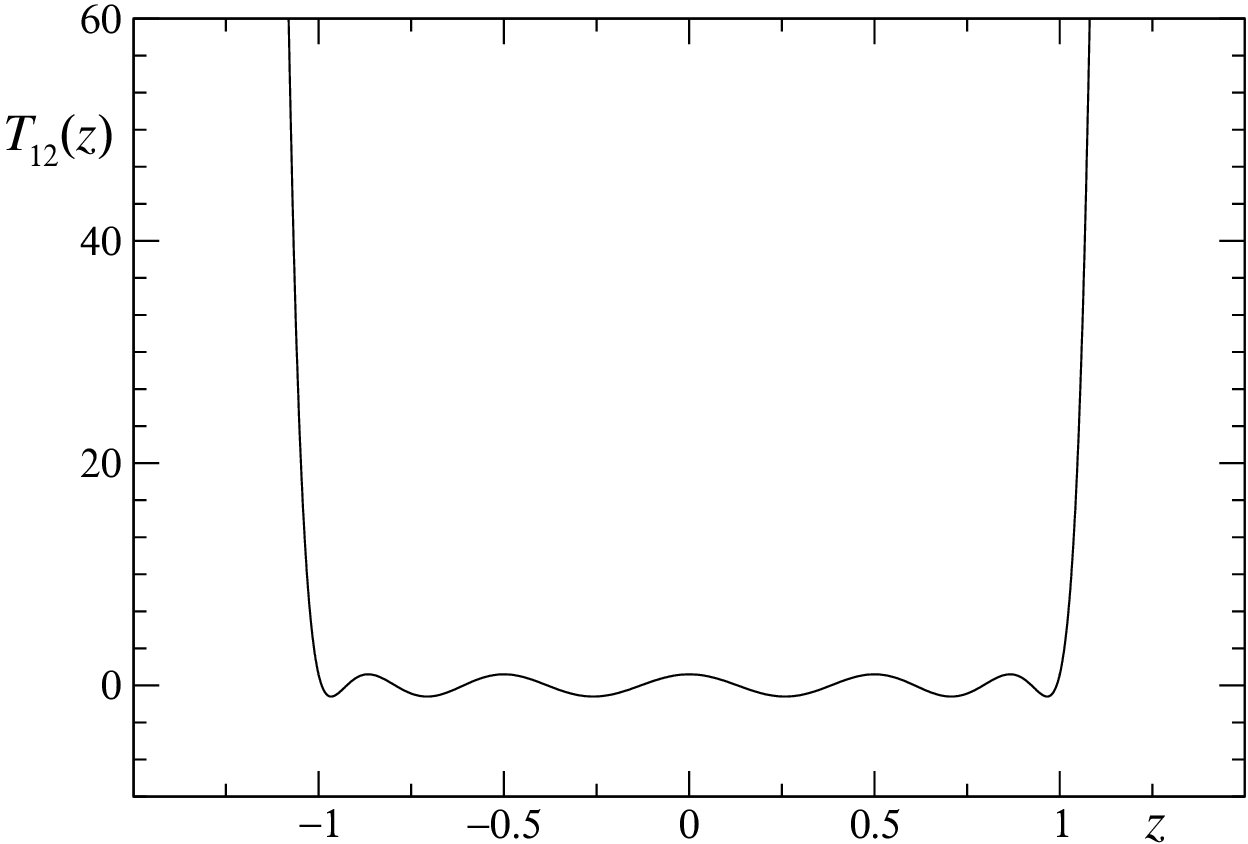}\hfill}
\vskip0.4cm
\figurecaption{%
Plot of the Chebyshev polynomial of order $12$, showing the bounded oscillatory 
behaviour in the interval $[-1,1]$ and the rapid increase
of the polynomial away from this range.
}
}
\endinsert

\subsection A.2 Chebyshev polynomials

For $\left|z\right|\leq1$ the Chebyshev polynomials
$T_0(z),T_1(z),\ldots$ are defined by
\equation{
  T_k(z)=\cos(k\theta), \qquad z=\cos\theta.
  \enum
}
They oscillate between $-1$ and $+1$ in this range and rapidly
increase or decrease to $\pm\infty$ when $|z|>1$,
depending on whether $k$ is even or odd (see fig.~7). Through the linear
transformation
\equation{
  z={2x-v-u\over v-u},
  \enum
}
the polynomials can be made to oscillate in an arbitrary 
interval $x\in[u,v]$ instead of the standard range $[-1,1]$. 

We may now replace $x$ by $A$ and choose
$[u,v]$ to contain the unwanted part of the spectrum of $A$. The other part
is then strongly enhanced when the polynomial is applied to 
a given quark field (see fig.~8).
Since the spectrum of the operator is not known beforehand, the interval 
bounds and the degree of the polynomial must be chosen adaptively in the 
course of the power iteration.

\topinsert
\vbox{
\vskip0.0cm
\epsfxsize=8.1cm\leftline{\hfill\epsfbox{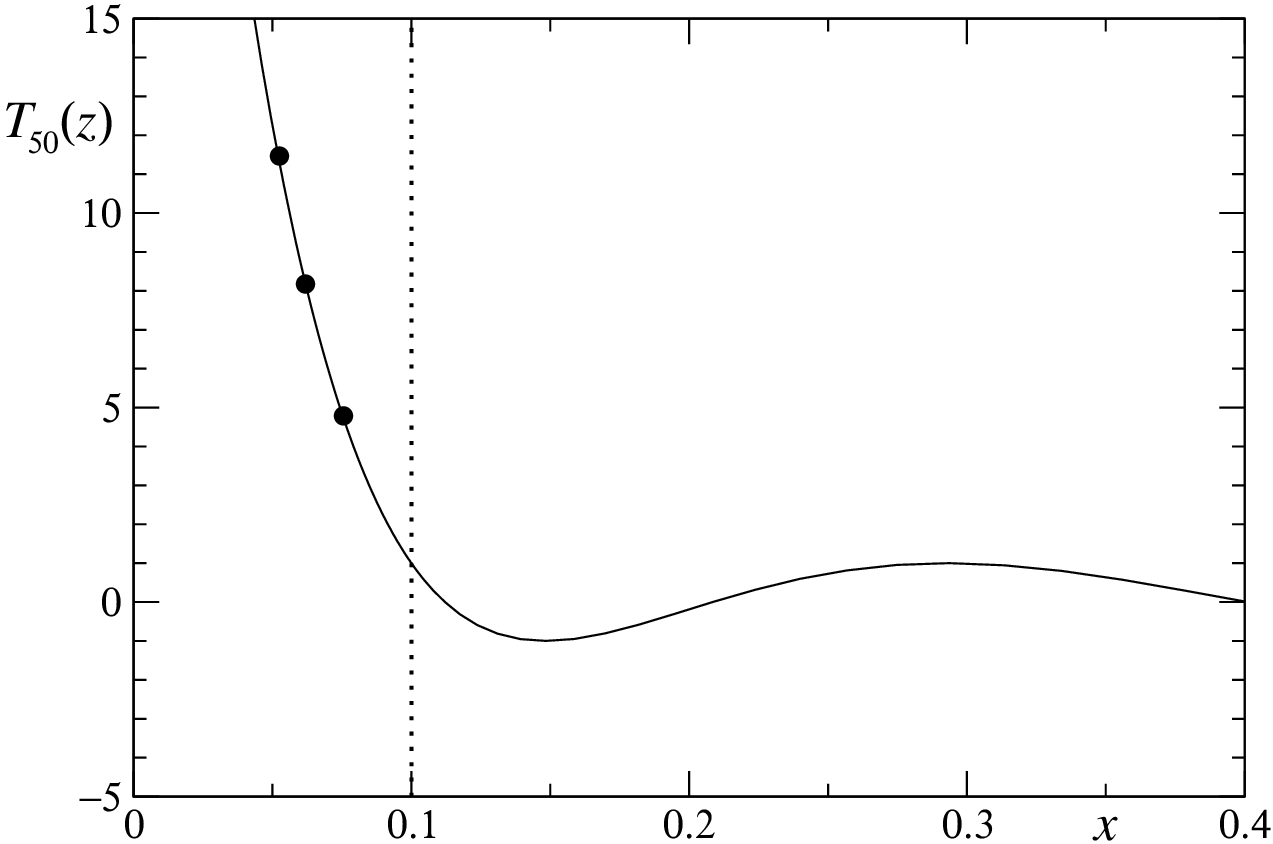}\hfill}
\vskip0.4cm
\figurecaption{%
Plot of the Chebyshev polynomial of order $50$, scaled to the interval
$[u,v]=[0.1,49]$. 
If the spectrum of $A$ is contained in this range
except for the few lowest eigenvalues (filled circles), 
the application of the polynomial to a given quark field
enhances the field components along the subspace spanned
by these modes.
}
}
\endinsert

A simple mathematical fact that will be used in this context is
summarized by

\proclaim
Lemma A.1. 
For any given even degree $k$, and real numbers $x,v,\gamma$
satisfying $x<v$ and $\gamma>1$,
there exists a unique value of the interval bound $u$ 
such that $x<u<v$ and $T_k(z)=\gamma$ [where $z$ is as in 
eq.~(A.3)].

\proof Since $\gamma$ is larger than $1$, there is one and only
one $\omega$ such that
\equation{
  \gamma=\cosh\omega, \qquad\omega>0.
  \enum
}
The inequality $x<u$, on the other hand, implies $z<-1$ and thus
\equation{
  T_k(z)=\cosh(k\nu),\qquad z=-\cosh\nu,
  \enum
}
for some $\nu>0$.
We then conclude that the equation $T_k(z)=\gamma$ has
a unique solution (respecting the specified constraints),
which is given by $\nu=\omega/k$. 

Having computed $z$ as a function of $k$ and $\gamma$, 
the interval bound 
\equation{
  u=x+(v-x)\tanh^2\Bigl({\omega\over2k}\Bigr)
  \enum
}
is obtained from eq.~(A.3). 
This shows that $u$ is uniquely calculable, and it is now also easy 
to verify that the expression (A.6) has all the required 
properties.

\subsection A.3 Chebyshev accelerated subspace iteration

The algorithm described in this subsection computes
increasingly accurate approximations to 
the $n$ lowest eigenvalues of $A$ and the associated eigenvectors.
It operates on a $d$-dimensional subspace of quark fields,
where $d$ is usually taken to be 
quite a bit larger than $n$. 

The computation proceeds iteratively, starting from a random
set of orthonormal quark fields $\psi_1,\ldots,\psi_d$.
In each cycle of the iteration, the fields are updated one 
by one so that at the end of the cycle the whole basis of fields
is replaced by a new one.
More precisely, a cycle consists of the following four steps:

\vskip1ex\noindent
1.~Choose a Chebyshev polynomial $T_k(z)$ with even degree and
an appropriate spectral range $[u,v]$,
taking the current estimates $\alpha_1,\ldots,\alpha_d$ of the eigenvalues
as input.

\vskip1ex\noindent
2.~Update the basis vectors $\psi_i$, $i=1,\ldots,d$, one after
another through
\equation{
   \chi=T_k(z)\psi_i,
   \qquad z={2A-v-u\over v-u},
   \enum
   \next{1ex}
   \phi=\chi-\sum_{j=1}^{i-1}(\psi_j,\chi)\psi_j,
   \enum
   \next{1ex}
   \psi_i=\phi/\|\phi\|.
   \enum
}
The second and third equations here simply implement the modified
Gram--Schmidt orthogonalization process. In particular, the new
basis is guaranteed to be orthonormal.

\vskip1ex\noindent
3.~Rotate the fields among themselves
so as to diagonalize the operator $A$ in the subspace spanned by 
them, i.e.~so that
\equation{
   (\psi_i,A\psi_j)=\delta_{ij}\alpha_i
   \quad\hbox{for all}\quad i,j=1,\ldots d,
   \enum
   \next{2ex}
   \alpha_1\leq\alpha_2\leq\ldots\leq\alpha_d,
   \enum
}
after the transformation. 

\vskip1ex\noindent
4.~Stop the algorithm if the approximate eigenvalues $\alpha_1,\ldots,\alpha_n$ 
and the associated eigenvectors
satisfy the chosen convergence criterion.

\vskip2ex\noindent
This description is somewhat schematic and needs to be made more precise.
It may not be obvious, for example, how to choose the polynomial in 
the first step, and there are various stopping criteria 
that may be applied.

\subsection A.4 Choice of the Chebyshev polynomial

Ideally the interval $[u,v]$ should 
contain the spectrum of $A$ except for the $d$ lowest eigenvalues.
In particular, the upper limit $v$ should be set to a value
slightly larger than $\|A\|$. If this number is not already known,
it may be calculated at the beginning of the subspace iteration 
using the ordinary power method.

To fix the lower bound $u$ of the spectral interval and the 
degree $k$ of the polynomial, a reasonable 
requirement is that
\equation{
  \left.T_k(z)\right|_{x=\alpha_1}=\gamma^2,
  \qquad 
  \left.T_k(z)\right|_{x=\alpha_d}=\gamma,
  \enum
}
where $\gamma>1$ is the desired enhancement factor for the 
low modes (see fig.~8).
In practice $\gamma=3$ appears to be a sensible choice,
but trying other values of $\gamma$ may be worth while.

Recalling lemma A.1, it is clear that the
conditions (A.12) determine both $u$ and $k$.
Explicitly, if we define $\omega$ and $\tilde\omega$ through
\equation{
   \cosh\omega=\gamma,\qquad\cosh\tilde\omega=\gamma^2,
   \enum
}
the conditions become
\equation{
  u=\alpha_d+(v-\alpha_d)\tanh^2\Bigl({\omega\over2k}\Bigr),
  \enum
  \next{1ex}
  \alpha_d+(v-\alpha_d)\tanh^2\Bigl({\omega\over2k}\Bigr)=
  \alpha_1+(v-\alpha_1)\tanh^2\Bigl({\tilde\omega\over2k}\Bigr).
  \enum
}
Normally $k$ is such that $2k\gg\tilde\omega$, and
the expansion at large $k$ of eq.~(A.15) then leads to 
\equation{
  k=\frac{1}{2}\left\{
  {\left(v-\alpha_1\right)\tilde\omega^2-\left(v-\alpha_d\right)\omega^2\over 
  \alpha_d-\alpha_1}
  \right\}^{1/2},
  \enum
}
while $u$ is determined by eq.~(A.14).
Evidently $k$ should be rounded to the closest
even integer, and one may also wish to impose lower and upper limits 
on $k$ at this point.

\subsection A.5 Stopping criterion

An upper bound on the deviation of the calculated eigenvalues 
$\alpha_i$ from the exact eigenvalues of $A$ 
may be obtained by computing the residues
\equation{
  \rho_i=(A-\alpha_i)\psi_i
  \quad (i=1,\ldots,n)
  \enum 
}
and the maximal eigenvalue $\epsilon^2$ of the $n\times n$ 
residual matrix 
\equation{
  R_{ij}=(\rho_i,\rho_j).
  \enum
}
A well-known lemma then asserts that
there are $n$ orthonormal eigenvectors of $A$
with eigenvalues $\hat{\alpha}_i$
such that $|\alpha_i-\hat{\alpha}_i|\leq\epsilon$
for all $i=1,\ldots,n$.

This convergence criterion is safe but can be inefficient 
if the set $\{\alpha_1,\ldots,\alpha_n\}$ of approximate eigenvalues divides
into well separated subsets of one or more
eigenvalues. The error bounds obtained from the
residual matrices associated to each subset of eigenvalues 
are then often quite a bit smaller than the bound
obtained from the total residual matrix.
Provided the subsets are 
indeed separated from one another, by a margin larger than 
the combined errors,
these tighter bounds are completely safe too.

\subsection A.6 Eigenvector locking

A fairly obvious property of the subspace iteration is that
the lower eigenvalues converge faster than the higher ones.
The algorithm can thus be accelerated somewhat, by
locking the subsets of eigenvalues and eigenvectors that
have already converged. Locking means that these eigenvectors
are not updated in the second step of each subspace iteration cycle 
and that, in the third step, the operator $A$ is diagonalized in the 
complementary subspace only. 

Another small acceleration is achieved by saving the last few
eigenvectors, say $\psi_{d-r+1},\ldots,\psi_d$, to some auxiliary
fields before they are updated in the second step. 
In the third step the saved fields may then be
included in the Ritz diagonalization, i.e.~$A$ is diagonalized
in a subspace of dimension $d+r$, but only the first $d$ 
eigenvectors are kept after the diagonalization.

\subsection A.7 Rounding errors

In order to avoid large rounding errors when
the Chebyshev polynomials are applied to the quark fields,
the Clenshaw recursion should be used [\ref{Recipes}].
Let $\psi$ be a given quark field and let us define
\equation{
  \chi_j=T_j(z)\psi, 
  \qquad z={2A-v-a\over v-u}.
  \enum
}
The computation then proceeds recursively according to 
\equation{
  \chi_0=\psi,\quad\chi_1=z\psi,
  \enum
  \next{2ex}
  \chi_{j+1}=2z\chi_j-\chi_{j-1},\qquad j=1,2,\ldots,
  \enum
}
until the desired degree $k$ is reached.

The use of 32-bit arithmetic in the subspace
iteration does not lead to uncontrolled rounding errors
as long as the degree $k$ of the 
Chebyshev polynomial is not too large.
In general the significance loss
in the Clenshaw recursion grows proportionally to $k$, and
degrees below 100 or 200 may therefore be safe.
On large lattices, however,
the lowest eigenvalues of $A$ tend to be closely spaced and many 
orders of magnitude smaller than the maximal eigenvalue.
Polynomials with significantly larger degrees
will be required under these conditions and the use
of double-precision arithmetic then becomes indispensable.

\beginbibliography


\bibitem{Wilson}
K. G. Wilson, 
Phys. Rev. D10 (1974) 2445


\bibitem{SW}
B. Sheikholeslami, R. Wohlert,
Nucl. Phys. B259 (1985) 572

\bibitem{OaImp}
M. L\"uscher, S. Sint, R. Sommer, P. Weisz,
Nucl. Phys. B478 (1996) 365


\bibitem{SchwarzAlgorithmII}
M. L\"uscher,
Comput. Phys. Commun. 165 (2005) 199

\bibitem{SchwarzReview}
M. L\"uscher, Lattice QCD with light Wilson quarks,
in: Proceedings of the 23rd International Symposium on Lattice Field Theory,
Dublin, 2005, eds. A. Irving, C. McNeile, C. Michael, PoS(LAT2005)002

\bibitem{NextPaper}
L. Del Debbio, L. Giusti, M. L\"uscher, R. Petronzio, N. Tantalo,
in preparation



\bibitem{GinspargWilson}
P. H. Ginsparg, K. G. Wilson,
Phys. Rev. D25 (1982) 2649


\bibitem{Hasenfratz}
P. Hasenfratz,
Nucl. Phys. B (Proc. Suppl.) 63 (1998) 53;
Nucl. Phys. B525 (1998) 401

\bibitem{HLN}
P. Hasenfratz, V. Laliena, F. Niedermayer,
Phys. Lett. B427 (1998) 125


\bibitem{NeubergerDirac}
H. Neuberger,
Phys. Lett. B417 (1998) 141;
{\it ibid.}\/ B427 (1998) 353


\bibitem{ExactChSy}
M. L\"uscher,
Phys. Lett. B428 (1998) 342


\bibitem{FerencReview}
F. Niedermayer,
Nucl. Phys. (Proc. Suppl.) 73 (1999) 105


\bibitem{HMC}
S. Duane, A. D. Kennedy, B. J. Pendleton, D. Roweth,
Phys. Lett. B195 (1987) 216


\bibitem{JansenSommer}
K. Jansen, R. Sommer (ALPHA collab.),
Nucl. Phys. B530 (1998) 185 [E: {\it ibid.} B643 (2002) 517]


\bibitem{SommerRadius}
R. Sommer,
Nucl. Phys. B411 (1994) 839


\bibitem{vanHemmen}
J. L. van Hemmen,
J. Phys. A: Math. Gen. 15 (1982) 3891


\bibitem{BanksCasher}
T. Banks, A. Casher,
Nucl. Phys. B169 (1980) 103


\bibitem{BernardGolterman}
C. W. Bernard, M. F. L. Golterman,
Phys. Rev. D49 (1994) 486

\bibitem{SharpeShoresh}
S. R. Sharpe, N. Shoresh,
Phys. Rev. D64 (2001) 114510


\bibitem{OsbornPQChPT}
J. C. Osborn, D. Toublan, J. J. M. Verbaarschot,
Nucl. Phys. B540 (1999) 317

\bibitem{DamgaardPQChPT}
P. H. Damgaard, J. C. Osborn, D. Toublan, J. J. M. Verbaarschot,
Nucl. Phys. B547 (1999) 305


\bibitem{BecirevicEtAl}
D. Be\'cirevi\'c et al. (SPQ$_{\rm CD}$R collab.),
Non-perturbatively renormalised light quark masses from a lattice simulation
with $N_{\rm f}=2$,
hep-lat/0510014 

\bibitem{DellaMorteEtAl}
M. Della Morte et al. (ALPHA collab.),
JHEP 0507 (2005) 007


\bibitem{EvaI}
B. Bunk, K. Jansen, M. L\"uscher, H. Simma,
Conjugate gradient algorithm to compute the low-lying
eigenvalues of the Dirac operator in lattice QCD,
ALPHA collaboration notes (September 1994)

\bibitem{EvaII}
T. Kalkreuter, H. Simma,
Comput. Phys. Commun. 93 (1996) 33


\bibitem{SubspaceI}
Y. Saad, Numerical methods for large eigenvalue problems
(Manchester University Press, Manchester, 1992); this book 
is out of print but can be downloaded from
{\tt http://www-users.cs.umn.edu/\~{}saad/books.html}


\bibitem{Recipes}
W. H. Press, S. A. Teukolsky, W. T. Vetterling, B. P. Flannery,
Numerical recipes in FORTRAN, 2nd ed. (Cambridge University Press,
Cambridge, 1992)

\endbibliography

\bye